\documentclass[%
reprint,
superscriptaddress,
nolongbibliography,
 amsmath,amssymb,
 aps,
pra,
]{revtex4-2}

\usepackage{graphicx}
\graphicspath{{./figures/}} 
\usepackage{dcolumn}
\usepackage{bm}
\usepackage{hyperref}
\usepackage[per-mode=symbol]{siunitx}
\DeclareSIUnit\gauss{G}
\usepackage{xfrac}
\usepackage{ragged2e}
\usepackage{multirow}
\usepackage[flushleft]{threeparttable}
\usepackage{xparse}

\DeclareDocumentCommand\vectorbold{ s m }{\IfBooleanTF{#1}{\boldsymbol{#2}}{\mathbf{#2}}} 
\DeclareDocumentCommand\vb{}{\vectorbold} 
\DeclareDocumentCommand\vectorunit{ s m }{\IfBooleanTF{#1}{\boldsymbol{\hat{#2}}}{\mathbf{\hat{#2}}}} 

\makeatletter
\def\@fnsymbol#1{\ensuremath{\ifcase#1\or \dagger\or *\or \ddagger\or
   \mathsection\or \mathparagraph\or \|\or **\or \dagger\dagger
   \or \ddagger\ddagger \else\@ctrerr\fi}}
\makeatother
\begin{document}

\preprint{SG-V2}

\title{Simulation of atom trajectories in the original Stern--Gerlach experiment}

\author{Faraz Mostafaeipour}
\thanks{These authors contributed equally.}
\affiliation{Division of Physics, Mathematics and Astronomy, California Institute of Technology, Pasadena, CA 91125, USA}
\affiliation{Caltech Optical Imaging Laboratory, Andrew and Peggy Cherng Department of Medical Engineering, Department of Electrical Engineering, California Institute of Technology, 1200 E. California Blvd., MC 138-78, Pasadena, CA 91125, USA}

\author{S.~S\"uleyman Kahraman}
\thanks{These authors contributed equally.}
\affiliation{Caltech Optical Imaging Laboratory, Andrew and Peggy Cherng Department of Medical Engineering, Department of Electrical Engineering, California Institute of Technology, 1200 E. California Blvd., MC 138-78, Pasadena, CA 91125, USA}

\author{Kelvin Titimbo}
\thanks{These authors contributed equally.}
\affiliation{Caltech Optical Imaging Laboratory, Andrew and Peggy Cherng Department of Medical Engineering, Department of Electrical Engineering, California Institute of Technology, 1200 E. California Blvd., MC 138-78, Pasadena, CA 91125, USA}

\author{Yixuan Tan}
\affiliation{Caltech Optical Imaging Laboratory, Andrew and Peggy Cherng Department of Medical Engineering, Department of Electrical Engineering, California Institute of Technology, 1200 E. California Blvd., MC 138-78, Pasadena, CA 91125, USA}

\author{Lihong V.~Wang}
\email[Email:]{lvw@caltech.edu}
\affiliation{Caltech Optical Imaging Laboratory, Andrew and Peggy Cherng Department of Medical Engineering, Department of Electrical Engineering, California Institute of Technology, 1200 E. California Blvd., MC 138-78, Pasadena, CA 91125, USA}

\date{\today}

\begin{abstract}
Following a comprehensive analysis of the historical literature, we model the geometry of the Stern--Gerlach experiment to numerically calculate the magnetic field using the finite-element method. 
Using this calculated field and Monte Carlo methods, the atomic translational dynamics are simulated to produce the well-known quantized end-pattern with matching dimensions. 
The finite-element method used provides the most accurate description of the Stern--Gerlach magnetic field and end-pattern in the literature, matching the historically reported values and figures.

\end{abstract}

\maketitle


\section{Introduction}\label{sec:intro}

In 1922, Otto Stern and Walther Gerlach reported what is now known as the Stern--Gerlach experiment (SGE), a fundamental milestone in the development of modern physics \cite{gers1922a,gers1922b,gers1922c}. 
Following Stern's preliminary work on the kinetic theory of gases and molecular beams \cite{stern_molecular_beam,stern_molecular_beam_addendum}, the SGE is a significant benchmark as it presented direct proof for angular momentum quantization \cite{sommerfeld1916theorie,debye1916quantenhypothese}, confirmed the electron intrinsic spin years later \cite{uhlenbeck_spinning_1926}, allowed for the selection of spin-polarized atoms, is the first measurement of atomic ground state properties without electronic excitation, and led to further research into entanglement, non-classical correlations, and the measurement problem \cite{einstein_ehrenfest_1922,schsl2016}.
As a result, the SGE has been widely used as an introduction and segue to quantum theory in contemporary textbooks \cite{sakurai_napolitano_2011,feynman_2011,messiah_2014}.
 
\begin{figure*}[t]
\centering
\includegraphics[width=0.95\linewidth]{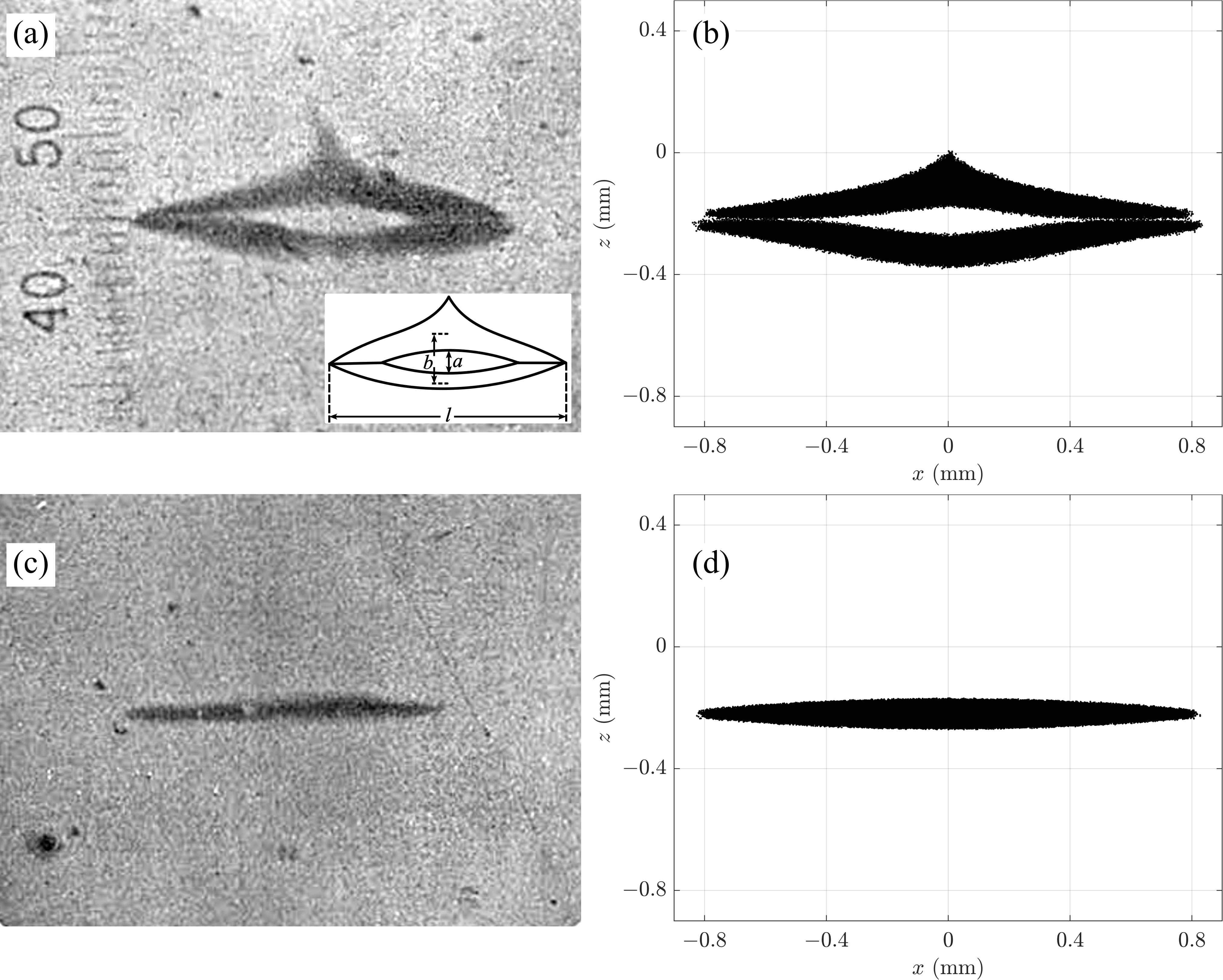}
\caption{Detected end-patterns for $\vb{B} \neq 0$ (top) and $\vb{B}=0$ (bottom). Panels (a) and (c) are the original patterns reported by Stern and Gerlach \cite{gers1922b}. Panels (b) and (d) correspond to the simulation outcomes. Inset of (a) is a representative illustration of the pattern displaying its length and width attributes reported in \cite{gers1922b}. Note, both patterns are to scale.}
\label{fig:EndPatternv3}
\end{figure*}

The three-paper series \cite{gers1922a,gers1922b,gers1922c} described the deflection of a beam of silver atoms entering an inhomogeneous magnetic field gradient.
The first of the series set out the preliminary framework, showing experimental proof of the silver atom's magnetic moment \cite{gers1922a}.
A few months later, in the second paper of the series, they published experimental verification of directional quantization, producing the well-known splitting patterns seen in Fig.~\ref{fig:EndPatternv3}(a) \cite{gers1922b}. 
In the third publication, they improved the precision of their measurements by addressing uncertainties related to the atomic beam's distance from the magnet and the strength of the magnetic field. 
Maintaining the same experimental apparatus, they measured ``the magnetic moment of the normal silver atom in the gaseous state [as] a Bohr magneton'' \cite{gers1922c,trageser_2022}. 

\begin{figure*}
\centering
\includegraphics[width=0.80\linewidth]{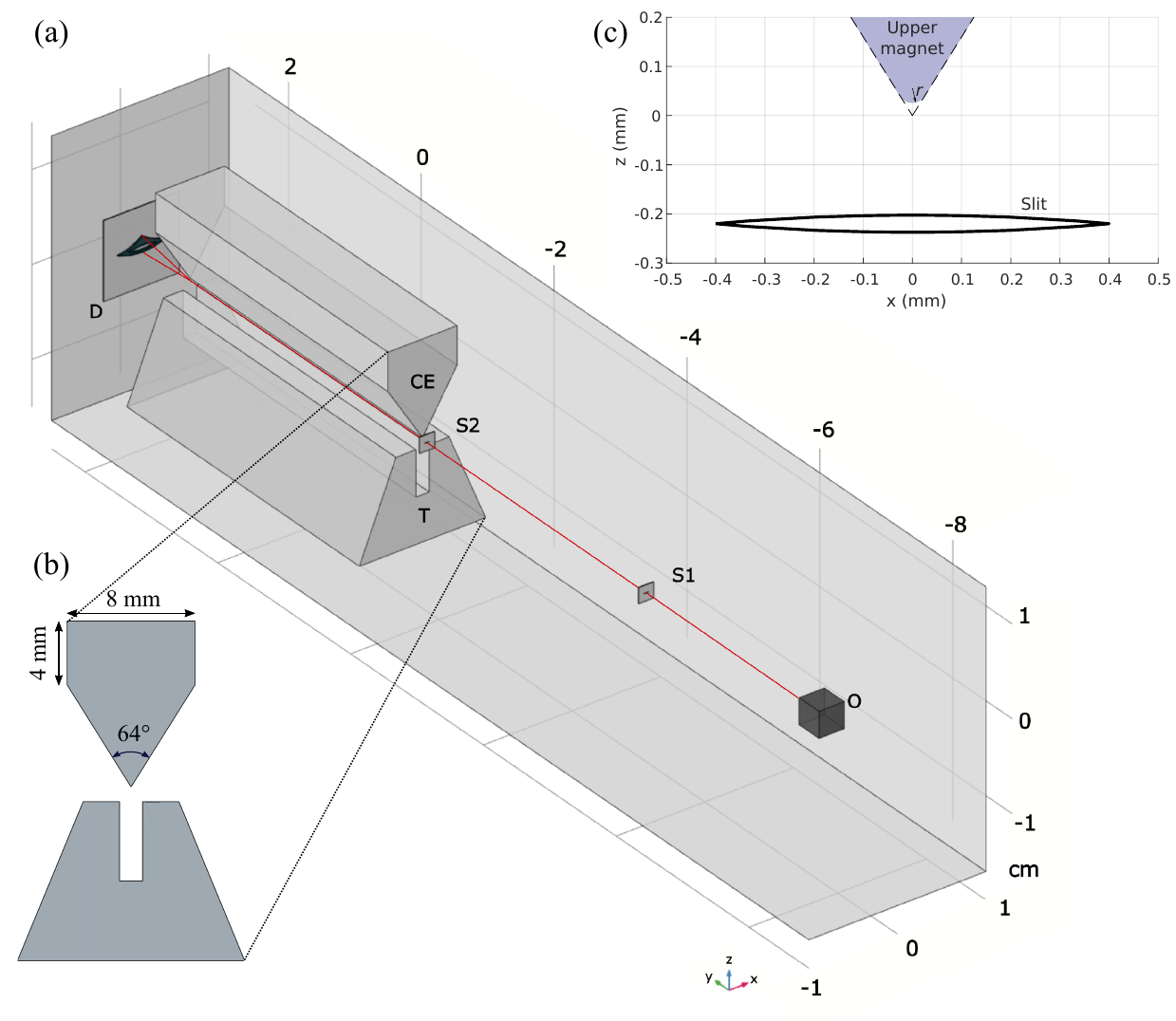}
\caption{Modelled Stern--Gerlach experiment. 
(a) Three-dimensional geometry of our model. O: oven. S1 and S2: slits. CE: magnet cutting edge. T: magnet trench. D: detector. 
(b) Cross-section of the magnets in the transverse plane. (c) Cross-section of Slit 2 (S2) with a slit height of \( h = \SI{0.035}{\mm} \) and a width of \( w = \SI{0.8}{\mm} \). The fillet with a radius of \SI{30}{\micro\metre} was introduced to smooth the CE tip.
}
\label{fig:geometry}
\end{figure*}

Fig.~\ref{fig:geometry} shows the model of the original SGE. 
Heated silver (Ag) atoms exit the oven (O) in the $y$-direction, travel through two collimating slits (S1 and S2), pass through an inhomogeneous magnetic field gradient generated by two ferromagnetic pole-pieces, and deposit on the end detector plate (D). 
The experiment shows the splitting of the beam into two components along the $z$ direction. 
The magnet consists of a blade-shaped pole and a furrowed pole called the cutting edge (CE) and trench (T), respectively (see Fig.~\ref{fig:geometry}(b)). 
Maximal splitting occurs at the center of the magnet, reducing as it moves away from the CE as seen in Fig.~\ref{fig:EndPatternv3}. 

In this work, we model the SGE magnetic field accurately with the COMSOL Multiphysics{${}^{\circledR}$} finite element analysis tool. 
In comparison, related literature studies on the SGE approximate the magnetic field analytically using Maxwell's equations, with a zero magnetic field along the atomic propagation axis and a spatially linear field dependence along the quantization axis \cite{raejm2022,ModernAnalysisSG,QM_SGE,home_pan_ali_majumdar_2007,alshm1982,hsu_berrondo_vanhuele_2011}. We perform a Monte Carlo simulation of the atomic trajectories traveling through this field. To our knowledge, this work provides the most accurate numerical replication of the original magnetic field and end-pattern of the historical SGE in comparison to the literature \cite{Wennerstrom2012, Rashkovskiy2022}.

\section{Simulation}
The field properties and the atomic trajectories are modelled based on historically accurate features and parameters from the original 1922 SGE papers and subsequent review studies \cite{gers1922a,gers1922b,gers1922c,trageser_2022,schsl2016}.

\subsection{Modeling the Magnetic field}

Finite-element analysis is used to simulate the magnetic field via COMSOL Multiphysics{${}^{\circledR}$} \cite{comsol}.
The three-dimensional geometry of the magnets has been drawn as closely as possible to the references. 
In COMSOL, \verb|extremely fine mesh| was selected to optimize the geometry and increase the field gradient resolution.
Furthermore, the cutting edge of the upper magnet was smoothed with a fillet to prevent numerical inaccuracies near the volume of interest.

\begin{figure}[t]
\includegraphics[width=0.95\linewidth]{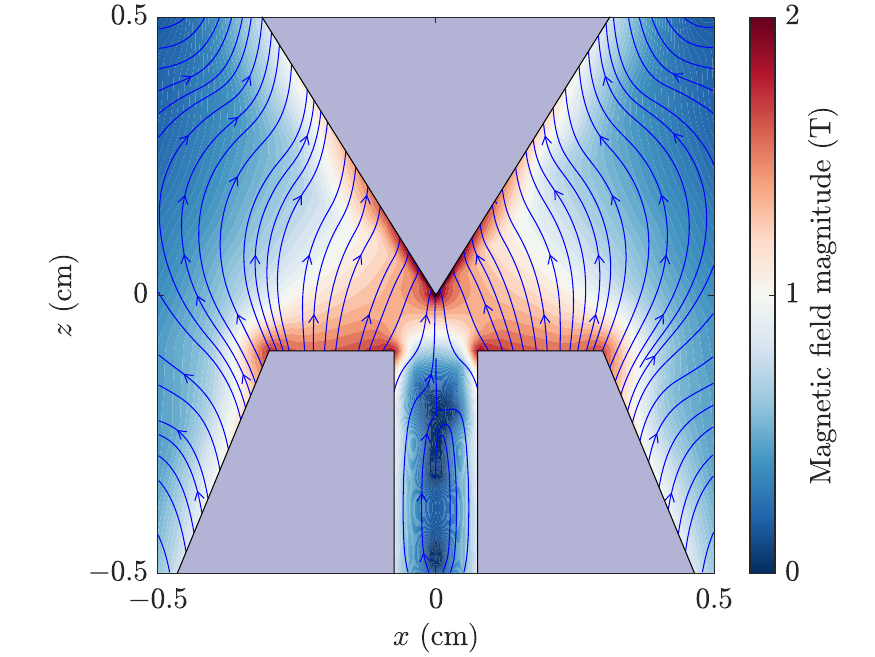}
\caption{Cross-section of the magnetic field on the $xz$ plane at $y = 0$. 
Both the upper and lower magnets are visible.
}
\label{fig:xz_cutplane}
\end{figure}

In Fig.~\ref{fig:geometry}, the reference frame's origin is centered at the middle of the magnets. The magnets span, \SI{3.5}{\cm} in length, is placed from $y=\qtyrange[range-units = single]{-1.75}{1.75}{\cm}$.
The magnets' cross-section is shown in Fig.~\ref{fig:geometry}(b) \& \ref{fig:xz_cutplane}. 
The trench of the lower magnet has a depth of \SI{5}{\milli\metre} and a width of \SI{1.5}{\milli\metre}. 
The CE tip of the upper magnet converges to the point \((x,z) = (0,0)\), features a wedge angle of \SI{64}{\degree}, and sits \SI{1}{\milli\metre} above the top of the trench.
The described cut-plane geometry is modelled after the experimental schematic presented in \cite{gers1922b}.

Non-smooth corners can introduce inaccuracies in finite element analysis. 
To mitigate these inaccuracies near the CE tip, a fillet with a radius of \SI{30}{\micro\metre} was introduced to smooth the corner, as shown in Fig.~\ref{fig:geometry}(c).
The precise shape of the CE tip is significant, as its proximity to the atomic trajectories means that any singularity in the magnetic field gradient can result in inaccuracies in the trajectories.

As shown in Fig.~\ref{fig:geometry}(a), the oven O is placed at $y = \SI{-7.65}{\cm}$ along the propagation direction, the slit S1 is positioned \SI{2.50}{\cm} away at $y = \SI{-5.15}{\cm}$, and the slit S2 is placed \SI{3.35}{\cm} away immediately in front of the SG magnets ($y = \SI{-1.80}{\cm}$). 
See reference \cite{gerlach4(2024),trageser_2022} for historical reference. 
The detector plate is placed immediately after the magnets. The slits O, S1, and S2 define the trajectory of the beam, which is centered at $(x,z) = (0,z_c)$ in the transverse plane. 
Hence, $z_c$ describes the vertical distance between the CE tip and the atomic beam at the entrance of the magnet in the absence of beam deflection before the magnets.
The scale and geometry of the simulation components are calculated in accordance with the literature \cite{gers1922b}.

\begin{figure}[t]
\includegraphics[width=0.95\linewidth]{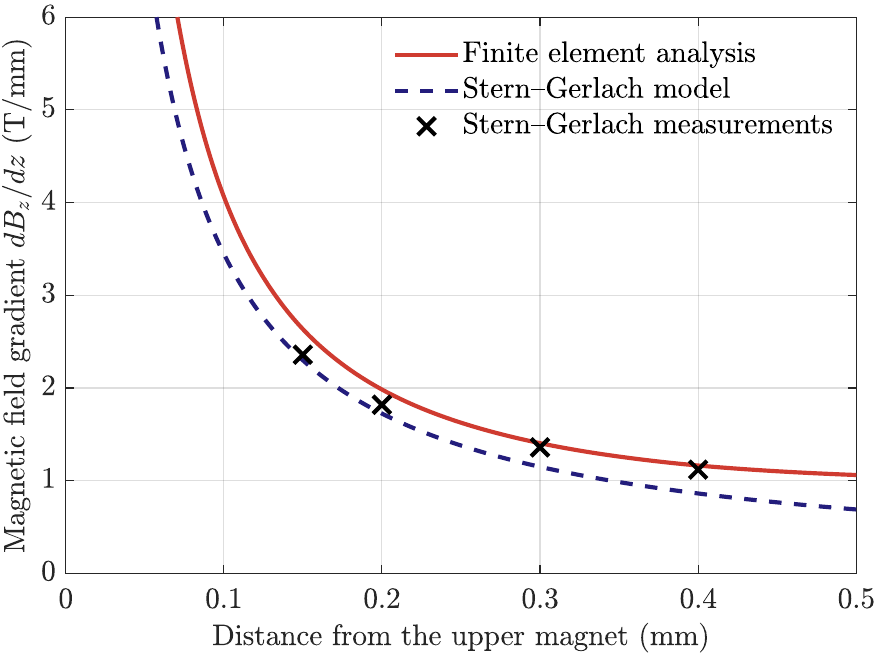}
\caption{(Red line) Gradient of the magnetic field along the \( z \) axis calculated using the finite element method. The same gradient is (blue dashed line) theoretically modelled and (black crosses) experimentally measured by Stern and Gerlach \cite{gers1922c}.}

\label{fig:gradient}
\end{figure}

Because there is no reference to the magnitude of the magnetic field that Stern and Gerlach specifically used for the reported pattern, we adjusted the field strength, $B$, such that $\frac{\partial B_{z}}{\partial z}$ as a function of $z$ matches the reported experimental values from the third 1922 SGE paper \cite{gers1922c}.
The theoretical model and experimental measurements of \( \frac{\partial B_{z}}{\partial z} \) presented in \cite{gers1922c} closely align with the COMSOL simulations shown in Fig.~\ref{fig:gradient}.

\begin{figure}
\centering
\includegraphics[width=0.95\linewidth]{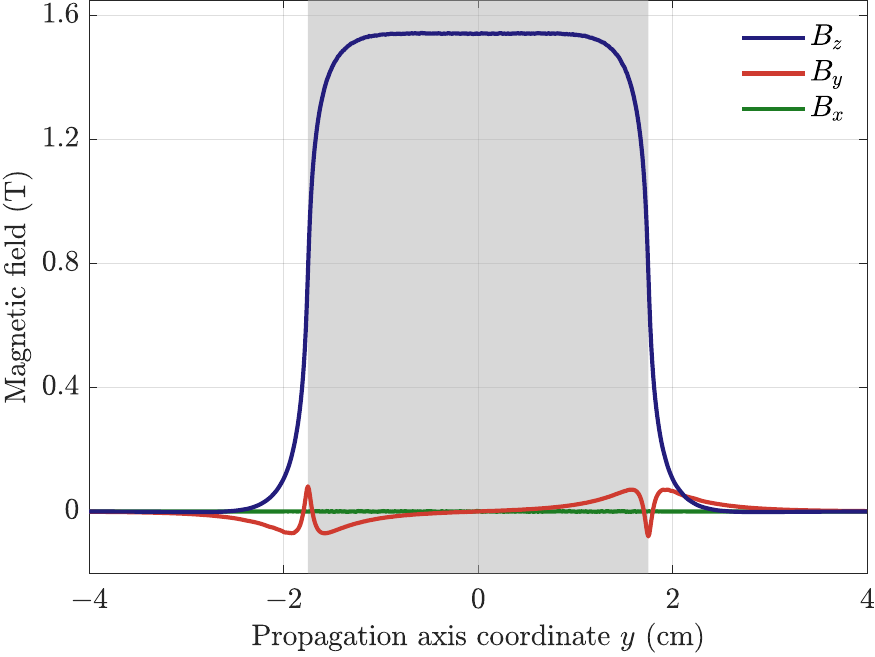}
\caption{Intensity of the magnetic field components: \( B_z \) (blue), \( B_y \) (red), and \( B_x \) (green) along the propagation axis \( (0,y,z_{c}) \). The shaded area indicates the location of the magnets. 
Due to symmetry, we have \( B_x = 0 \). 
Because $\nabla\cdot\vb{B}=0$, \( B_y \) shows small fluctuations near the ends of the magnets.}

\label{fig:Bfield}
\end{figure}

The magnetic field cross-section centered at $(x,z) = (0, z_{c} = -\SI{0.22}{\milli\metre})$ along the atomic propagation axis ($y$) is shown in Fig.~\ref{fig:Bfield}.

\subsection{Atomic trajectories}

A Monte Carlo simulation is calculated where $N = \num{e7}$ Ag atoms ($M = \SI{107.868}{\dalton}$) are ejected from the oven's circular aperture with an area of \SI{1}{\mm^{2}}, populated based on a random radial Gaussian distribution centered at $(x,z) = (0,z_c = \SI{-0.22}{\milli\metre})$. 
The atomic translational dynamics are numerically calculated using Euler's method based on the classical equation of motion for the neutral atoms in the experiment:
\begin{align}
    \label{eq:muz} \frac{\mathrm{d}^{2}\vb{r}}{\mathrm{d}t^2} = -\dfrac{1}{M} \bm{\nabla} \left(- \vb{\bm{\mu}}_{e} \cdot \vb{B}(\vb{r}) \right )  \ 
\end{align}
where $t$ is time, $\vb{r}$ is the position of the atom, $\vb{\bm{\mu}}_{e}$ is the magnetic moment of the electron, $\vb{B}(\vb{r})$ is the magnetic field at coordinate $\vb{r}$. 
Our simulation code is available online for reference \cite{GITHUB}. 

For simplicity, we assume that the electron magnetic moments are statistically evenly split to parallel or anti-parallel to the field immediately out of the oven; no magnetic moment is assumed to be oriented at an oblique angle.
This assumption is equivalent to solving two independent trajectory equations for each eigenstate for a spin of $\sfrac{1}{2}$ \cite{hsu_berrondo_vanhuele_2011}. 
Various descriptions are available in the literature, but solving two independent equations for each state is widely implemented \cite{dev2015,schsl2016,raejm2022}.
For comparison, atoms with classical unquantized magnetic moments (sampled from an isotropic distribution) are simulated, yielding an expected continuous end-pattern (see Fig.~\ref{fig:classicalpattern} in Appendix).

The atoms' initial angular distribution of velocities are sampled from within a solid cone spreading outward in the positive $y$-direction with a vertex angle selected such that it fills S1.
The speeds are uniformly sampled in the range $|\vec{v}|=[625,750]\unit{\metre\per\second}$. 
Details about the velocity selection are further outlined in the discussion section.
From the oven, the atoms travel to S1's circular aperture with an area of \SI{3e-3}{\mm^2}.
The atoms passing through S1 travel further towards S2.
The original SGE papers describe S2 as a slit-shaped aperture with a length of \SI{0.8}{\mm} and a width of 0.03$-$\SI{0.04}{\mm}.
In the absence of a magnetic field gradient, it is expected that the shape of the end-pattern on the detector plate is a scaled projection of the S2 shape.
Based on this, we modelled the slit as an eye-shaped aperture to match the zero-field ($B=0$) end-pattern with correct historical length and width \cite{gers1922b}. 
The contour of the aperture is defined as the intersection of the two parabolae: 
\begin{equation}
    f_{z,\pm}(x) = \pm \frac{2h}{w^2}x^2 + z_{c} \mp \frac{h}{2} , \qquad -\sfrac{w}{2} \leq x \leq \sfrac{w}{2} ,
\end{equation}
as shown in Fig.~\ref{fig:geometry}(c).
From S2, the atoms traverse through the magnet and hit the detector plate placed immediately after the magnet at $y = \SI{1.8}{\cm}$ (see Fig.~\ref{fig:geometry}(a) and references \cite{gerlach4(2024),trageser_2022} for reference). 
As previously mentioned, the oven, slits, and therefore, the atomic beam are centered at $(x,z)=(0, z_c=-0.22\ \text{mm})$. 
The simulation does not consider reflections off the magnets and disables particles that interact with the slits, magnets, and surrounding environment. 

\section{Discussion}
As seen from Fig.~\ref{fig:gradient}, the simulated magnetic field gradients agree with the experimentally reported values in the literature.
Additionally, the selection of $z_c = \SI{-0.22}{\mm}$ is such that $B_z$, as shown in Fig.~\ref{fig:Bfield}, is close to the reported magnetic field value of around $\sim \SI{1.65}{\tesla}$ \cite{gerlach4(2024)} and the final atomic distribution on the screen resembles the original pattern. 
From the simulations, we find that the depth of the trench does not alter the magnetic field enough to perturb the end-patterns. 
From our literature review, this work is the most comprehensive description of the magnetic field used in the original SGE experiment.

As shown in Fig.~\ref{fig:EndPatternv3}, Stern and Gerlach measure $a_{\text{exp}} = \SI{0.11}{\mm}$ as the minimum splitting, $b_{\text{exp}} = \SI{0.2}{\mm}$ as the average splitting, and $L_{\text{exp}} = \SI{1.1}{\mm}$ as the pattern length \cite{gers1922b}.
Given the historical description of the SGE, several variables, such as $z_c$ and propagation velocities, remain historically unknown and can be tuned within an appropriate range to match the end-pattern in \cite{gers1922b}.

As the SGE was an experimental consequence of Stern's preliminary work on molecular velocity measurements, the methodology for producing an atomic silver beam is similar. 
The approach is based on heating a silver wire to its melting point, causing it to evaporate under vacuum, thus creating an isotropic propagation of atoms, which is collimated into a beam.
Placing a cold metal plate along the beam path causes the vapor to condensate on the detector \cite{trageser_2022}.
In the SGE, they used a silver-loaded clay furnace for more stability over time, but the principle of emission is the same \cite{gerlach4(2024)}. 
By using the modified Maxwell's velocity distribution for atomic beams (where atoms with higher velocities have a higher probability of entering the atomic beam) for a given temperature, the root-mean-squared velocity, $v_{\text{rms}} = \sqrt{4 k_b T/m}$, is found \cite{stern_molecular_beam_addendum,stern_molecular_beam,ramsey_2005}. 
The melting temperature of silver is \SI{1234.9}{\kelvin} and is the minimum temperature needed to create the atomic beam \cite{rumble_bruno_doa_2022}. 
The original SGE reports a temperature of \SI{1323}{\kelvin} \cite{gerlach4(2024)}. 

From our simulation, we found that the maximum and minimum speeds selected from the distribution are correlated to the minimum splitting and average splitting values, respectively. 
The speed determines the time of flight of the entire path and the interaction time with the magnets. 
Selecting $v_{\text{max}}=\SI{750}{\metre\per\second}$ and $v_{\text{min}}=\SI{625}{\metre\per\second}$ yields a simulated $a_{\text{sim}}= \SI{0.11}{\mm}$ and $b_{\text{sim}}= \SI{0.2}{\mm}$, which is in good accordance with a Maxwellian speed distribution with a temperature of \SI{1323}{\kelvin}. 
In reference to Fig.~\ref{fig:EndPatternv3}(c) \& (d), the SGE reports the width of the pattern at its widest point in zero-field conditions as \SI{0.10}{\mm}, which is consistent with our simulation \cite{gers1922b}. 

\begin{figure}[ht]
\centering
\includegraphics[width=0.98\linewidth]{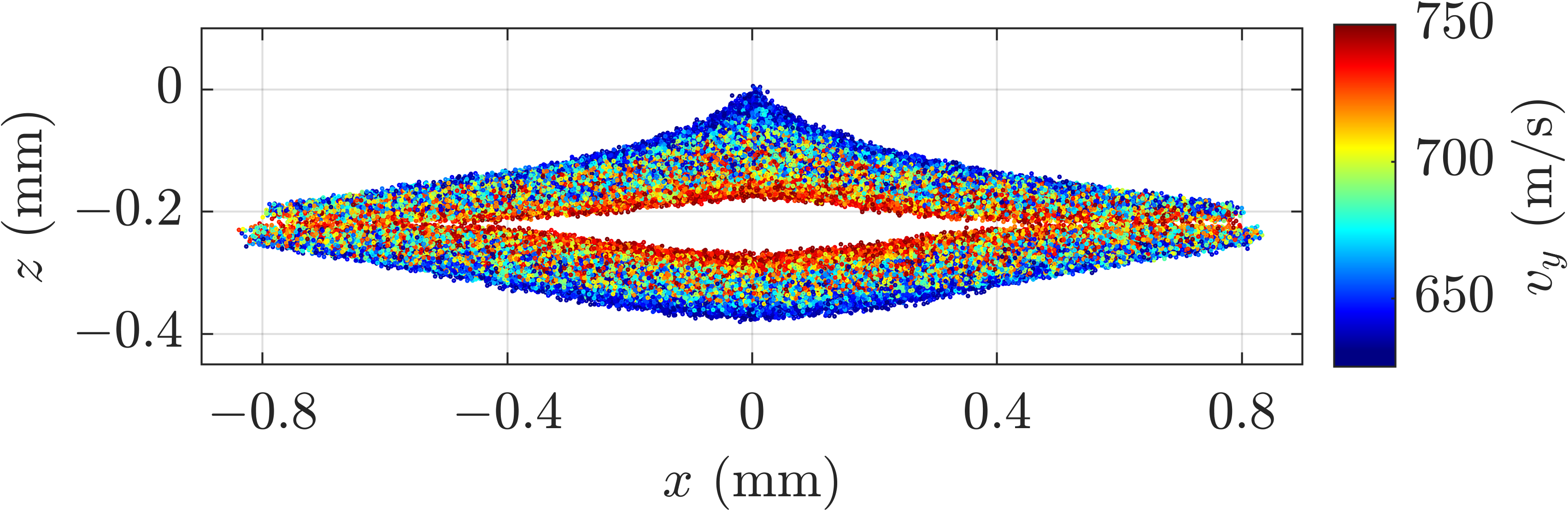}
\caption{End-pattern for an ensemble of atoms colored according to their initial speeds in the $y$ direction, $v_y$.}
\label{fig:vz_map}
\end{figure}

For a given propagation speed, there is an associated pattern with a given splitting distance. 
Lower speeds yield longer flight time to interact with the magnets, thus splitting more. 
As a result, the final shape is an accumulation of these sub-patterns following the speed distribution. 
The end-patterns associated with $v_{\text{max}}$ and $v_{\text{min}}$ are shown in Fig.~\ref{fig:vz_map}. 
Distinct color bands are seen for the given speeds, where the overlap is explained due to the initial angular distribution of velocities.  
In the large $N$ limit, the end-pattern on the detector is saturated as all the velocities will be sampled. 
Therefore, we have selected a uniform distribution of speeds between $v_{max}$ and $v_{\text{min}}$. 
This assumption is reasonable as Stern and Gerlach noted the difficulty in developing the detector film as the deposited silver was too thin to see with the unaided eye, reporting long irradiation times (eight hours) and saturating the film \cite{gers1922b, friedrich_herschbach_2003}.

\begin{figure}[ht]
\centering
\includegraphics[width=0.98\linewidth]{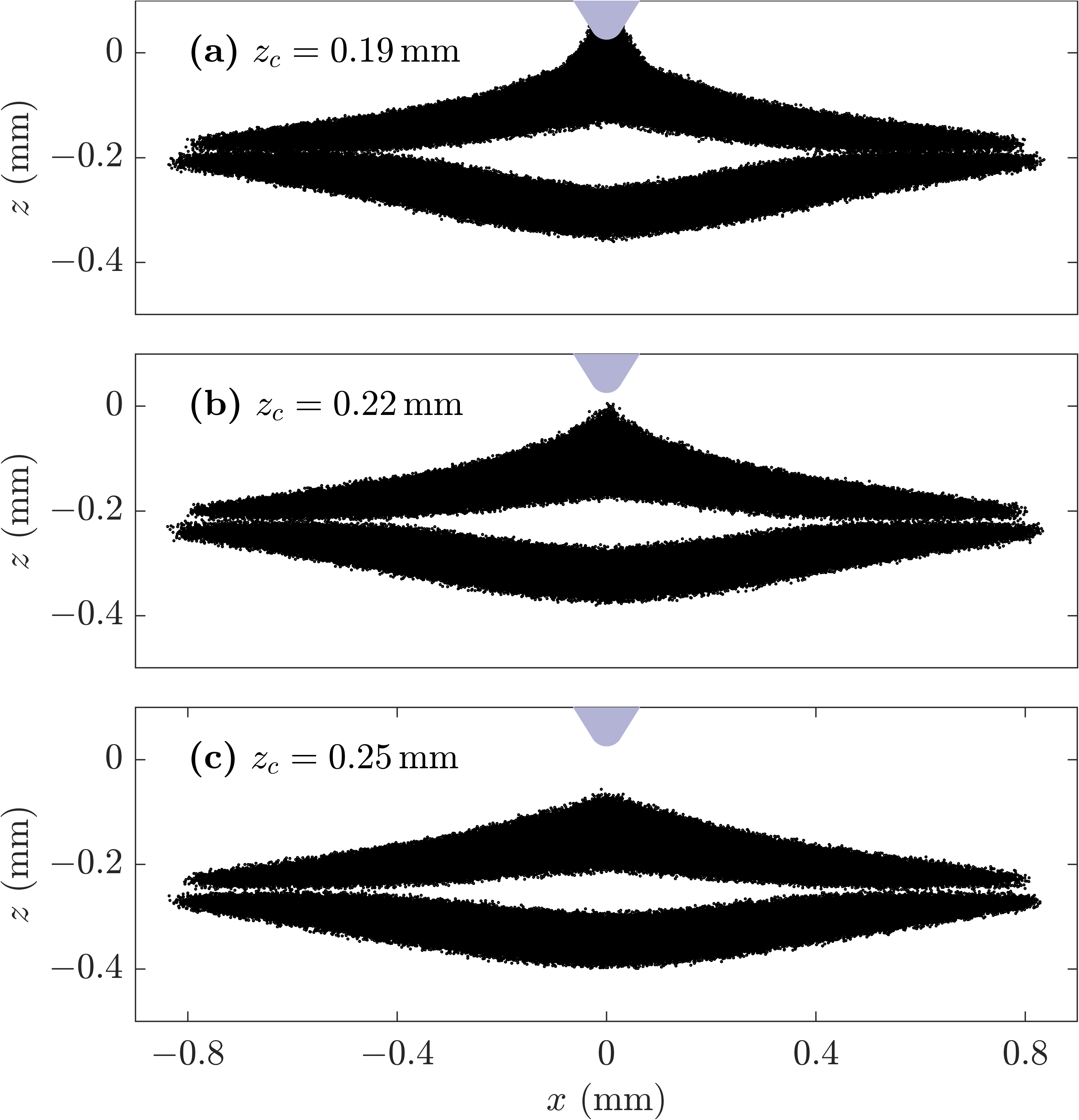}
\caption{End-patterns for different distances between the slits and the upper magnet, $z_c$. Atoms traveling closer to the CE tip of the upper magnet experience stronger field gradients.}
\label{fig:zc}
\end{figure}

In Fig.~\ref{fig:EndPatternv3}, the simulated length is measured as $L_{\text{sim}} \approx \SI{1.7}{\mm}$ in contrast to $L_{\text{exp}}$. 
The length of the pattern is correlated to the initial angular distribution of velocities and divergence from a collimated beam. 
In our simulation, the angular distribution and the divergence, governed by O, S1, and S2 geometries, determine the length of the pattern. 
However, the discrepancy can be explained by a historical depiction of the full SGE apparatus, showing a series of uncharacterized slits and apertures between O and S1 (see references \cite{gerlach4(2024),trageser_2022}).  
These additional apertures reduce the angular deviations and shorten the length of the pattern. 
In support of this assumption, we placed a third slit to test the effect of additional collimators and observed a reduced length matching the experimental length while maintaining the shape of the pattern. 
Because additional slits are not detailed in the SGE papers, we omitted them for the final pattern.

\section{Conclusion}
In summary, we modelled the SGE based on the properties, dimensions, and geometries from the historical papers and subsequent review studies. 
The field properties are calculated using the COMSOL Multiphysics{${}^{\circledR}$} finite element analysis tool. 
We obtained a simulated magnetic field and gradient agreeing with experimentally reported values. 
Using this field, our Monte Carlo method simulated the atomic trajectories to obtain an end-pattern. 
To our knowledge, our simulated magnetic field and end-pattern are the most accurate numerical descriptions of the SGE.

\begin{acknowledgments}
We would like to thank D.~C.~Garrett, Z.~He, and A.~Bengtsson for their insights when discussing this work.
This project has been made possible in part by grant number 2020-225832 from the Chan Zuckerberg Initiative DAF, an advised fund of the Silicon Valley Community Foundation.
\end{acknowledgments}

\appendix
\renewcommand\thefigure{\thesection.\arabic{figure}} 

\section{Classical end-pattern}
\setcounter{figure}{0} 

\begin{figure}[ht]
\centering
\includegraphics[width=0.98\linewidth]{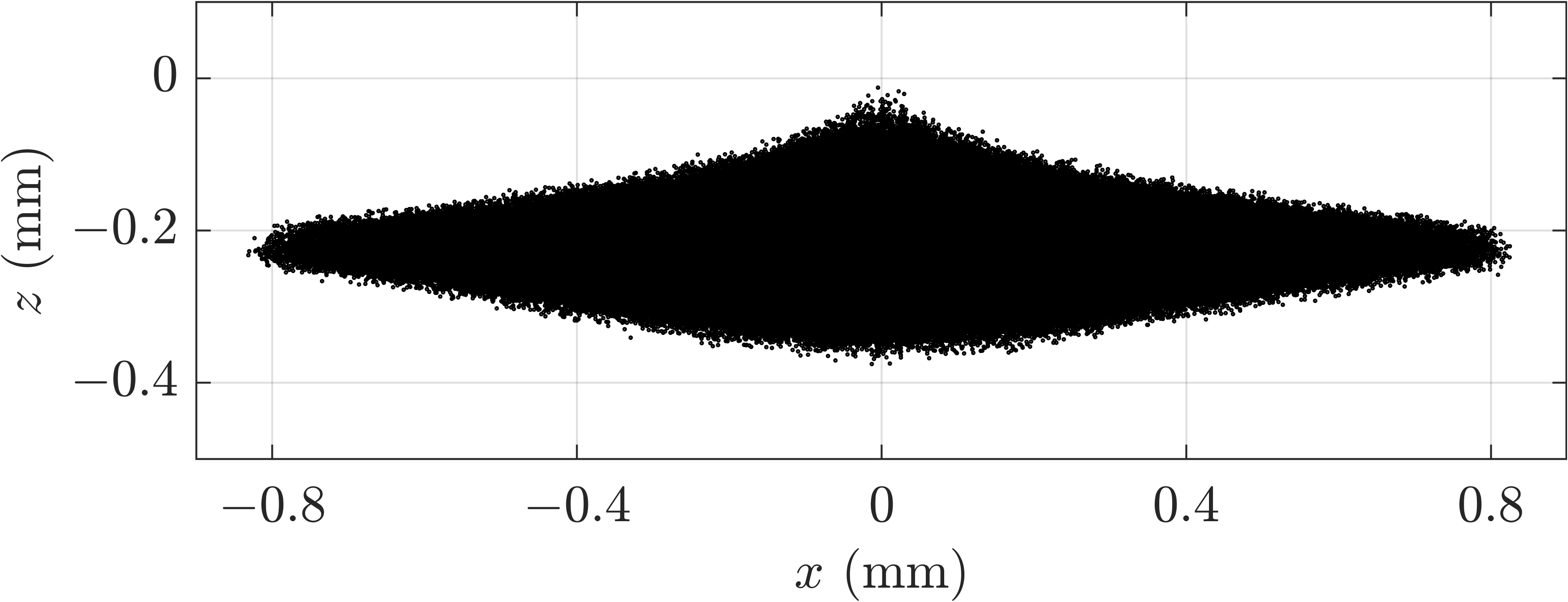}
\caption{Classical end-pattern for an ensemble of atoms with a non-quantized magnetic moment sampled from an isotropic distribution.}
\label{fig:classicalpattern}
\end{figure}

When the magnetic moment of the valence electron of the silver atom is not considered quantized, the expected observation on the detection plate is shown in Fig.~\ref{fig:classicalpattern}.
No splitting is observed.

\bibliography{apssamp}

\end{document}